\documentclass[a4paper]{jpconf}
\usepackage{graphicx}
\begin{document}
\title{Search for Single and Pair-Production of Dijet Resonances with the CMS Detector}

\begin{center}
{\bf (Proceedings submitted to the Kruger 2012 Conference) }
\end{center}

\author{Kai Yi for the CMS Collaboration}

\address{Physics and Astronomy Department, University of Iowa, Iowa, USA}

\ead{yik@fnal.gov}

\begin{abstract}

Searches for new physics in  the single and paired   
dijet mass spectrum are performed using data 
collected by the CMS experiment at the LHC at a collision energy 
of $\sqrt{s}$=7 or $\sqrt{s}$=8 TeV. No evidence for new physics is 
found and upper limits are set for various models. 
At 95\% confidence level, a string resonance in the single dijet 
spectrum is excluded for masses between 1 and 4.7 TeV and, for the 
first time, a coloron in the paired dijet spectrum is excluded 
for masses between 250 and 740 GeV.

\end{abstract}

\section{Introduction}
Many beyond-the-standard-model models predict the existence 
of new massive strongly produced objects that decay into 
quarks ($q$) and gluons ($g$). These states may show up as 
narrow resonances in the leading (highest $p_T$) jet spectrum.
We consider the following models: 
string resonance ($qg$, $q\bar{q}$ 
and $gg$)~\cite{Anchordoqui:2008di,Cullen:2000ef}; 
$E_6$ diquark ($\bar{q}\bar{q}$ and $qq$) predicted from grand unified 
theory based on the $E_6$ gauge group~\cite{ref_diquark};
excited quarks  $q^*$ ($qg$) expected from composite quark model~\cite{ref_qstar,Baur:1989kv}; 
axial vector particles called axigluons $A$ ($q\bar{q}$)~\cite{ref_axi,Chivukula:2011ng}; 
color-octet vector $C$ ($q\bar{q}$)~\cite{ref_coloron}; 
color-octet scalar $S_8$ ($gg$)~\cite{Han:2010rf};
new gauge bosons $W^\prime$ and $Z^\prime$ from new gauge 
symmetries ($q\bar{q}$)~\cite{ref_gauge};  
massive gravitons $G$ ($q\bar{q}$ and $gg$) from the Randall-Sundrum (RS) 
model of extra dimensions~\cite{ref_rsg}.
The color-octet vector is predicted by the flavor-universal coloron 
model, which embeds the SU(3) symmetry of QCD in a larger gauge 
group; the color-octet scalar is included in many dynamical 
electroweak symmetry breaking models, such as technicolor.

In addition to singly produced particles, many models predict the pair 
production of massive colored particles~\cite{paridijet}.
The searches of new narrow resonance in the leading dijet mass spectrum 
is not optimized for pair-produced particles for the following reason:
the probability that one of the particles decays into the two leading dijet 
is only about 10\%. A dedicated search is needed for pair-produced 
particle.  
We use coloron  production as our benchmark 
model ($gg\rightarrow CC\rightarrow q\bar{q}q\bar{q}$). 
We also consider the possibility that colorons decay into  
color-octet scalars ($S_8$). Furthermore, we consider 
an R-Parity violating SUSY model~\cite{susymodel}
in which pair-produced top squark (stops) 
each decay to $q\bar{q}$ as a third possibility.

\section{Jets Reconstruction at CMS}

Jets  are reconstructed using a particle flow
technique~\cite{pflow} at CMS~\cite{cmsdetector}. 
The CMS particle-flow algorithm combines calorimeter 
information with reconstructed tracks to identify   
photons, leptons, and both neutral and charged hadrons within jets.
The particle-flow objects
serve as input for jet reconstruction, performed using the
anti-$k_T$ algorithm~\cite{ak5} with a cone radius $R=\sqrt{(\Delta\eta)^2 + (\Delta\phi)^2}$ of 0.5 or 0.7.
Jet energy scale corrections are derived from Monte Carlo (MC)
simulation ~\cite{jec} as well as data.

\section{Dijet Resonances Searches}

We search for narrow resonances in the dijet mass spectrum 
using the two  leading jets in the data.
The data sample is collected with   multijet high-level triggers  
based on $H_T$, the scalar sum of the transverse momenta of all jets in 
the event with $p_T > $ 40 GeV  
and   $|\eta|<2.5$. 
The $H_T$ threshold  are 650 GeV and 750 GeV, depending on the 
running period. 
The trigger efficiency is measured from the data to be larger than 99.9\% for 
dijet masses reconstructed offline above 890 GeV.
Jets that pass the standard jet quality requirements are combined 
into ``wide jets''~\cite{widejet} if their separation 
$R=\sqrt{(\Delta\eta)^2+(\Delta\phi)^2}\le1.1$. The combining is done 
before determining the leading dijet mass spectrum.

\begin{figure}[hbtp]
  \begin{center}
    \includegraphics[width=0.6\linewidth]{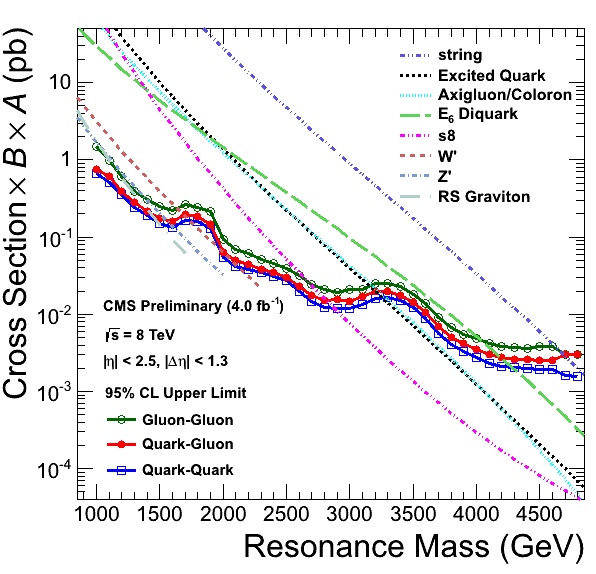}
    \caption{The observed 95\% CL upper limits from the high-mass 
analysis on $\sigma\times B\times A$ for dijet 
resonances decaying through gluon-gluon (open circles), 
quark-gluon (solid circles), and quark-quark (open boxes) channels 
compared with theoretical predictions for string 
resonances~\cite{Anchordoqui:2008di,Cullen:2000ef}, 
$\mbox{E}_6$ diquarks~\cite{ref_diquark},
excited quarks~\cite{ref_qstar,Baur:1989kv}, 
axigluons~\cite{ref_axi,Chivukula:2011ng},
colorons~\cite{ref_coloron}, $S_8$ resonances~\cite{Han:2010rf}, 
new gauge bosons $\mbox{W}^{\prime}$ and $\mbox{Z}^{\prime}$~\cite{ref_gauge},
and Randall-Sundrum gravitons~\cite{ref_rsg}.
}
    \label{figLimit}
  \end{center}
\end{figure}

The data used in this study correspond to an integrated luminosity of 
4.0 fb$^{-1}$ at a collision energy of $\sqrt{s}$=8 TeV~\cite{cmsdijet}.
The dijet mass background shape is modeled by 
a smooth parameterization.  
The dijet resonance shapes for  $qq$, $qg$ and $gg$ partons are dominated  
by experimental resolution and are obtained from full simulation.
No significant excess is observed.
The upper limits at 95\% CL set on the cross section times branching 
ratio times acceptance of centrally ($|\Delta \eta|<1.3$ and $|\eta|<2.5$) 
produced dijet mass resonances
for ($qq$, $qg$, $gg$)  are shown in 
Fig. \ref{figLimit}.
We set lower limits on the mass of string resonances, 
excited quarks, axigluons, colorons, 
$S_8$ resonances, $\mbox{E}_{6}$ diquarks, 
W$^{\prime}$  and Z$^{\prime}$ bosons, and RS gravitons in 
the 1--4.7 TeV range 
as listed in Table~\ref{exclusionLimits_fat}.
Many of the values in the table represent significant improvements on 
exclusion limits from previous dijet mass searches.

\begin{table}[ht]
  \centering
  \normalsize
  \begin{tabular}{|c|c|c|c|}
  \hline
    Model & Final State & Obs. Mass Excl. & Exp. Mass Excl. \\
          &             & [TeV]     & [TeV] \\
    \hline
String Resonance (S) & qg  & [1.0, 4.69]  & [1.0,4.64] \\
Excited Quark (Q*) & qg  & [1.0, 3.19]  & [1.0,3.43] \\
$E_{6}$ Diquark (D) & qq  & [1.0, 4.28]  & [1.0,4.12] \\
Axigluon (A)/Coloron (C) & $\mbox{q}\bar{\mbox{q}}$  & [1.0, 3.28]  & [1.0,3.55] \\
$S_8$ Resonance ($S_8$) & gg  & [1.0, 2.66]  & [1.0,2.53] \\
W' Boson (W') & $\mbox{q}\bar{\mbox{q}}$  & [1.0, 1.74]  & [1.0,1.92] \\
              &                           & [1.97, 2.12]  &            \\
Z' Boson (Z') & $\mbox{q}\bar{\mbox{q}}$  & [1.0, 1.60]  & [1.0,1.50] \\
RS Graviton (RSG) & $\mbox{q}\bar{\mbox{q}}$+gg  & [1.0, 1.36]  & [1.0,1.20] \\
\hline
  \end{tabular}
  \caption{Observed and expected 95\% CL exclusions on the mass of various resonances. Systematic uncertainties are included.}
\label{exclusionLimits_fat}
\end{table}

\section{ Searches for Pair-produced Dijet Resonances in a Four Jet Final State  }

We search for new narrow resonances in the paired dijet mass spectrum. 
The data sample is collected with triggers 
that require the presence of at least four 
jets, based on information from the calorimeters. Each jet must 
have $|\eta|<3.0$ 
and $p_T>70$  or 80 GeV, depending on the run period. 
The trigger is 99.5\% efficient for events with our selection--four 
leading  jets, each with a $p_T$ exceeding 
110 GeV with $|\eta|<2.5$.
The two jets in every possible pairs is required to have 
a separation on $\Delta R_{jj}=\sqrt{(\Delta \eta)^2+(\Delta \phi)^2}\ge0.7$ 
to ensure a negligible overlap among the jets. 
There are three possible dijet pairs from the leading four 
jets, and each dijet pair is required 
to have $\Delta{m}<0.15m_{avg}$, where $\Delta{m}$ is the mass 
difference between the two dijets and $m_{avg}$ is their average 
mass. Thus the difference is required to be less than three times 
the dijet mass resolution (4.5\%).

%to have $\Delta m < 0.15 m_{avg}$, 
%which is approximately three times the dijet mass resolution of 4.5\%, 
%where $\Delta m$ is the mass difference between the two dijets and 
%$m_{avg}$ is their average mass. 

We require $\Delta>25$ GeV  for each dijet, 
where $\Delta=\Sigma_{i=1,2}(p_T)_i-m_{avg}$ is the scalar sum 
of the transverse momenta of the two jets in the dijet and the average pair 
mass in the event.  
The $\Delta$ requirement removes a broad structure around 300 GeV 
from QCD events and ensures a smoothly falling dijet mass background 
shape. It also reduces 
the QCD background  by more than an order of magnitude 
while retaining approximately 25\% of the signal.

\begin{figure}
\centering
\includegraphics[width=0.6\linewidth]{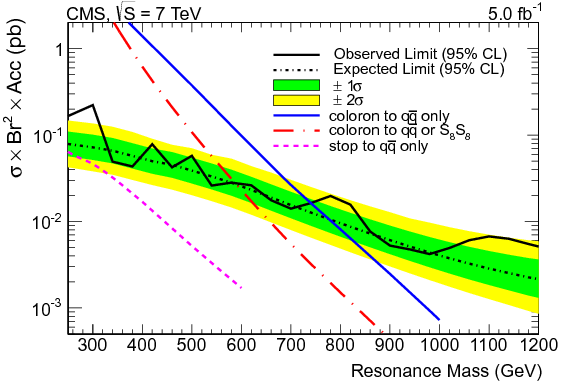}
\caption{
The observed and expected 95\% CL limits on the product of the 
resonance pair production cross section, the square of the 
branching fraction to dijets, and the detector acceptance, given 
by the solid and dot-dashed black curves, respectively. 
The shaded regions indicate the 1$\sigma$ and 2$\sigma$ bands 
around the expected limits. Predictions of a coloron model and a 
SUSY model are also shown.
}
\label{fig:pairdijetlimit}
\end{figure}

The data used in this study correspond to an integrated luminosity of 
5.0 fb$^{-1}$ at a collision energy of $\sqrt{s}$=7 
TeV~\cite{Chatrchyan:2013izb}.
No evidence of new physics  is found in the data. 
The expected and observed 95\% CL upper limits as a function of mass 
are presented in Fig.~\ref{fig:pairdijetlimit}. 
For the first time, at 95\% CL, the pair production of colorons is 
excluded for coloron 
masses between 250 and 740 GeV assuming that a coloron decays 100\% 
into $q\bar{q}$, or between 250 and 580 GeV assuming that coloron 
decays into $q\bar{q}$ compete with decays into $S_8S_8$.
With the current integrated luminosity there is some sensitivity to 
a SUSY model for pair-produced stops.

\section{Conclusion}

We performed searches for new physics in the singly-produced 
and pair-produced dijet mass spectrum.  
No sign for new physics is found.  Upper limits are set for various models 
and significant parameters space are excluded.
We exclude the string resonance between 1 TeV and 4.7 TeV in the 
searches from the single dijet mass spectrum, and exclude the coloron 
with a mass between 250 GeV and 740 GeV for the first time 
in the searches from the paired dijet mass spectrum.

\end{document}